# Application of Distributed Fiber Optic Strain Sensors to LMQXFA Cold Mass Welding

M. Baldini*, S. Krave*, R. Bossert, S. Feher, T. Strauss, and A. Vouris.

*Abstract*— **The future High Luminosity upgrade of the Large Hadron Collider (HL-LHC) at CERN will include the low-beta inner triplets (Q1, Q2a/b, Q3) for two LHC insertion regions. The Q1, Q3 components consist of eight 10 m-long LMQXFA cryo-assemblies fabricated by the HL-LHC Accelerator Upgrade Project. Each LMQXFA Cold mass contains two Nb3Sn magnets connected in series. A stainless-steel shell is welded around the two magnets before the insertion into the cryostat. There is a limit on how much coil preload increase induced by the shell welding is allowed. Distributed Rayleigh backscattering fiber optics sensors were used for the first time to obtain a strain map over a wide area of a Nb3Sn magnet cold mass shell. Data were collected during welding of the first LMQXFA cold mass and the results confirm that the increase of the coil pole azimuthal pre-stress after welding do not exceed requirements.**

*Index Terms*— **Accelerator magnets, HL-LHC, Nb₃Sn, Optical fiber sensors, Strain measurement**

## INTRODUCTION

The USA Accelerator Upgrade for the HiLumi-LHC (US-HL-LHC AUP) project is fabricating ten Q1/Q3 cold masses for the interaction regions of the High Luminosity Large Hadron Collider (HL-LHC) [1]. The HL-LHC interaction region magnet triplet consists of three optical elements: Q1, Q2, and Q3. Q1/Q3 cryo-assemblies contain two MQXFA quadrupole magnets.

The LMQXFA cold mass is the He pressure vessel assembly containing two 4.2 m long Nb₃Sn MQXFA superconducting magnets. A stainless-steel (SS) shell is used as a pressure vessel [2, 3]. The two half shells are simultaneously welded together on both sides of the MQXFA magnets. The interference between the stainless-steel shell and the magnet Aluminum (Al) shells need to be controlled. Due to the fragile nature of the Nb₃Sn conductor, there is a limit on how much pre-stress generated by the SS shell welding, the MQXFA magnet can tolerate. In the MQXFA Magnet Interface Specification [3] it is specified how much stress is acceptable to prevent damaging the coil [4]. A maximum of 3.2 MPa azimuthal coil pre-load increase is allowed in the pole. This value has been determined considering that the difference between the inner circumference of the SS shell and the outer circumference of the Al shell of the magnet needs to be ≥-0.2 mm. This corresponds to a SS vessel average azimuthal stress of around 15.5 MPa (i.e., 21 MPa for the outer circumference of the SS shell and 10 MPa for the inner circumference of the same shell) [5]. In case of weld repairs, a maximum of 8 MPa coil pre-load increase in the pole and a SS vessel stress of 40 MPa is locally allowed [5].

The allowed SS shell stress has been calculated using finite element analysis from the difference between the stainless shell inner diameter circumference and the magnet outer diameter circumference. To prevent coil high preload, the design incorporates a 2 mm thick, and 15 mm wide shim tacked to the inside surface of the cold mass shell and located at the top and bottom center and running through the full length of the magnets. This shim allows the stainless shell to bend at the gaps versus stretching during the longitudinal welding and thus to reduces the preload on the shell resulting in a lower preload on the coils [5]. A detailed discussion can be found in References [5-9]. After the welding process is completed the weld shrinkage and shell stretching (comparing measured length values before and after welding) are measured. Actual strain measurements are also performed. This manuscript is focused on presenting and discussing the strain measurements data collected during the welding process for the first LMQXFA cold mass.

A high-definition distributed fiber sensor was used to measure the strain variations during welding and verify that the coil preload increase meets the requirements [5]. The working principle of the optical sensor exploits the Rayleigh backscattering due to fiber natural imperfections. Those defects are scattering centers for elastic Rayleigh scattering. Some fraction of the scattering events results in backscattering and are used to calculate the spectral shift as a function of length on the fiber. The optical system required to measure the spectral shift is based on Swept Wavelength Interferometry (SWI) [10]. The elastic Rayleigh spectrum can be measured, and it is expected to be modified because of strain and temperature variations [11].

Those fibers have been demonstrated to be a very promising tools for quench detection in high temperature superconducting magnets. For example, fibers were integrated into a REBCO conductor architecture and demonstrated strain sensing

This work was supported by the U.S. Department of Energy, Office of Science, Office of High Energy Physics, through the US LHC Accelerator Upgrade Project (AUP).

M. Baldini, S. Krave, R. Bossert, S. Feher, T. Strauss, A. Vouris are with Fermi National Accelerator Laboratory, Batavia, IL 60510 USA, (e-mail: mbaldini@fnal.gov).

capabilities as well as thermal perturbation detection and quench localization with higher spatial resolution than voltage taps [12-13]. In this manuscript, a novel approach to measure strain on Nb3Sn magnet was developed using distributed fiber optics and tested for the first time. Therefore, the goals of this study are twofold: verify that the strain increase on the SS-shell during the welding of the LMQXFA cold mass meets requirements and demonstrate the novel use of distributed optical strain sensors to generate strain maps of Nb3Sn superconducting magnets.

## I. FIBER INSTALLATION ON THE LMQXFA SS-SHELL

A 10 m long high-definition fiber optic sensor from LUNA INNOVATION has been employed for the welding test. The diameter is 155 μm and, the fiber is coated with polyimide.

A grid of 700 x 400 mm was designed according to the fiber length as shown on the top left panel of Fig. 1. A one-to-one model was used to draw the pattern on the upper quadrant of the SS shell (see top panel of Fig. 1). The fiber was installed transferring the model drawing onto the shell and connecting the dots on the drawing (Fig. 1 top and bottom left panels). The fiber was then glued with epoxy as displayed on the bottom right panel of Fig. 1. The SS shell radius is 310 mm which corresponds to a 480 mm arc length. Therefore, the fiber sensor grid covers most of the shell quadrant. Its longitudinal position is close to the center of the cold mass. The welding seam is 100 mm below the grid as shown on the bottom right panel of Fig. 1.

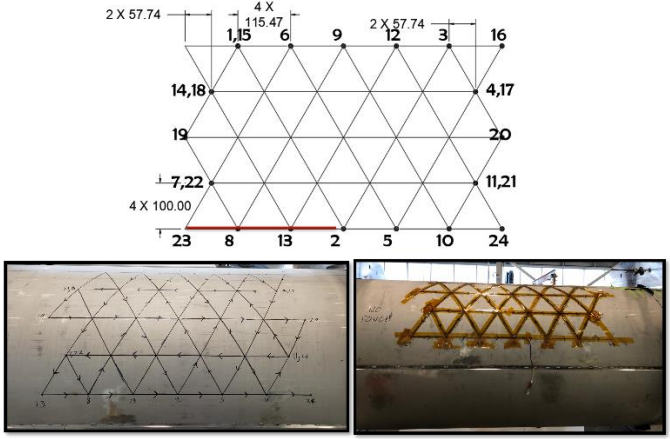

Fig.1: Fiber grid design (upper panel ), the red line marks the section from where the data plotted in Fig 3 are taken; grid drawing on the shell (bottom left ); fiber installation and gluing (bottom right)

Specific positions along the fiber were identified touching the fiber with a Q-tip before collecting the data as suggested by LUNA procedures. Those positions are the ones labeled by increasing numbers in the grid design (see the upper panel of Fig. 1) and allow to identify in which part of the fiber a particular value of strain is measured.

Those fiber sensors can acquire data with several spatial resolutions. During the welding of the first LMQXFA cold mass a 0.65 mm spatial pitch was used together with a sample rate of 0.52 sample/seconds. This is equivalent to having 15000 measuring points along the 10 m optical sensor. A tare was collected before the welding process and used to normalize the final strain values. The spectrometer allows to get real-time data as a function of fiber length and time. On the top panel of Fig. 2 an example of a real time data set is reported. The raw data consist of strain variations measured along the entire length of the fiber sensor at a specific time. An easier to visualize plot of the same data set is made using the fiber grid to produce a strain color map as displayed in Fig. 2b.

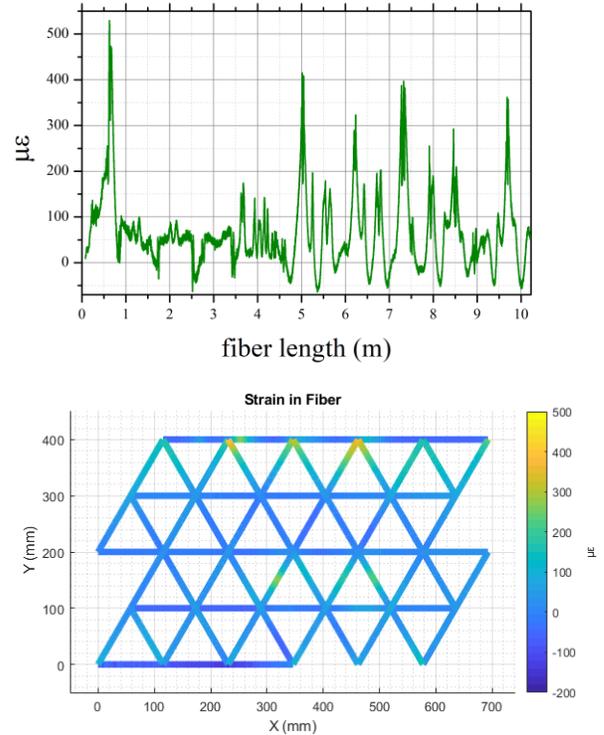

Fig.2: (a): Example of raw strain data taken at a specific point in time; (b): strain variation measured along the fiber grid

The actual welding of the LMQXFA cold mass took place at the y= -100 mm location starting from the left side and moving toward the right. An initial strain wave was observed from the welder passing the gage area. As the shell cooled down, the strain redistributed around the skin. This is evident looking at the strain data measured along the straight sections of the grid that is closer to the welder passage (see Fig. 3). Each data set correspond to a spatial position along the fiber line connecting point 23 to 2 on the grid displayed in Fig. 1. As soon as the welder got closer to the grid area, strain was observed to increase, to reach a maximum and then to decrease as the area thermalized. Indeed, the presence of a strain peak is consistent with the raise of temperature due to welding.

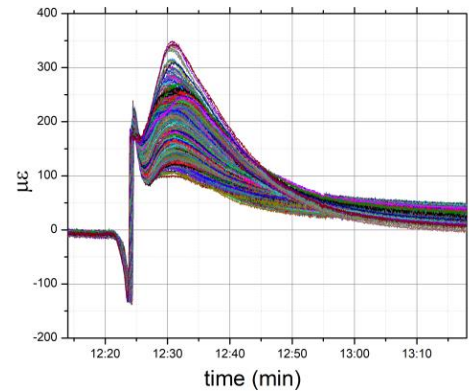

Fig.3: Data collected over a longitudinal section of the grid underlined with a red line in Fig. 1. The section is around 33 cm long. The axial strain variation measured as a function of time is displayed here. Each curve corresponds to a data point taken on the selected section of the sensor. Strain measurements are taken every 0.65 mm; there are around 513 data points.

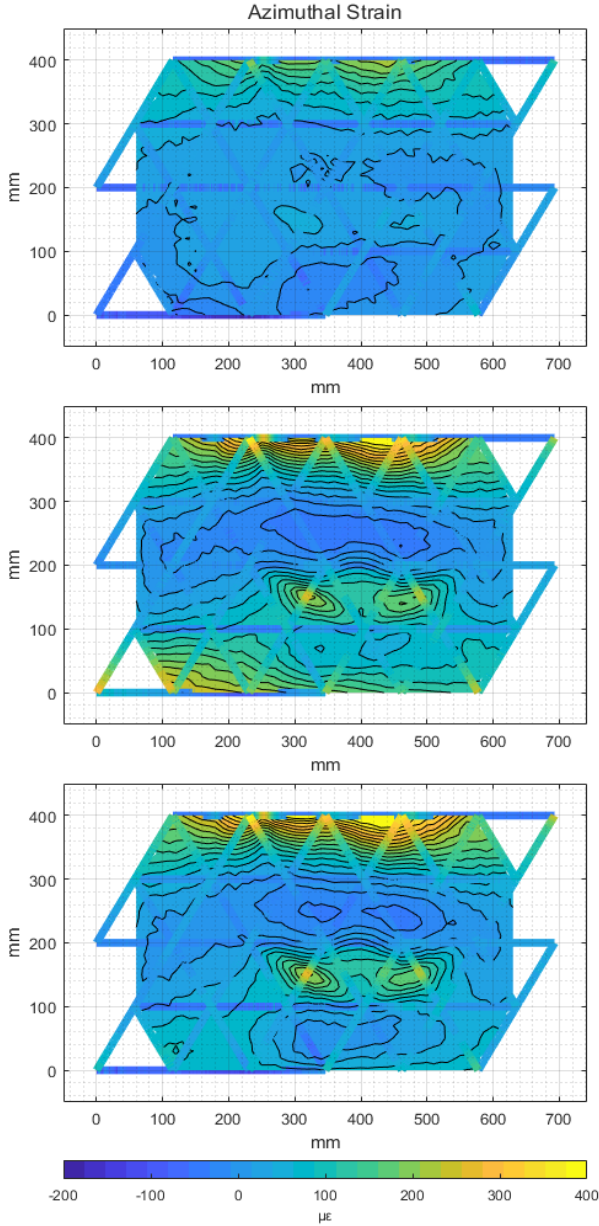

Fig.4: Azimuthal strain map snapshots of the SS vessel during the second welding passage.

## II. STRAIN MAP

The fiber grid was designed with isometric triangles to exploit one of the special cases of Strain Rosette Layouts [14, 15]

Similarly, to what it is done with strain gauges, the 60 degrees strain Rosette equations were used to calculate the strain in each direction for each fiber crossing on the grid. In this case the x strain direction is the 0 orientation of the fiber. The principal strain was obtained together with strain along the x (axial) and y (azimuthal) directions. Each direction had been interpolated over a rectangular grid, generating a strain map of the entire area covered by the grid. It is good to note that this interpolation is valid only for the area bounded by points in which strain values exist for each direction. There was an initial concern about the strain measured at crossover regions, but the data did not show any evidence of discontinuities. The fiber sensor grid covers a 100000 times larger area than a single strain gauge, making it a very powerful and reliable tool.

A strain map movie of the entire welding process was obtained using this method. In Fig.4 three strain map snapshots are displayed (each contour plot corresponds to a 20 με variation). The top one provides a strain representation at the very beginning of the welding, the second one a representation of the strain in the middle of the welding process and the bottom one a strain representation at the end of the welding process. The temperature effect due to the welder passage can be easily iden-tified since the strain variation observed in the bottom part of the grid is much more preeminent in the middle of the welding process, and it is almost gone at the end of the welding. On the other hand, strain appears to be consistently maximum on the upper part of the grid which is closer to the top of the SS shell. This reflects the fact that the MQXFA magnets inside the shell do not have a perfectly circular shape, but something closer to a diamond-like shape, making the upper area of the magnet the first one entering into contact with the SS-shell.

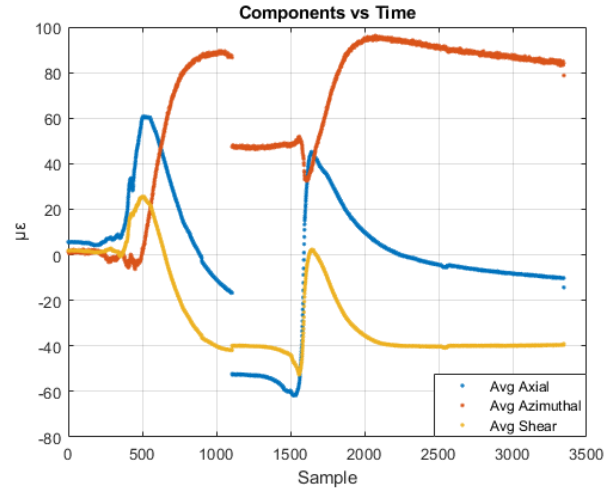

Fig.5: Azimuthal, axial and shear average strain measured during the first and second welding passage.

Before presenting the quantitative results, it is important to underline that those strain data reflect both the bending and the stretching effects on the SS-shell. The only way to decouple bending from stretching effects is to have an equivalent grid installed in the inner circumference of the shell. On the other hand, the shim layer installed between the SS-shell and the Al cylinder of the MQXFA magnet favors the SS shell bending at the gaps versus stretching during the longitudinal welding. For this reason, it is reasonable to believe that the strain values measured during this test are mostly due to bending and thus they are overall overestimated.

The average azimuthal (Y-direction on the strain map displayed in Fig.4), axial (X-direction) and shear strain was calculated using all the points observed in the strain map. Three welding passages were performed in a span of three days. The tare strain values collected at the beginning of the process were used to normalize the datasets.

In Fig. 5 the average strain vs recorded data points is displayed for the first two welding passages. The fiber was broken before disconnecting it from the spectrometer after the last welding passage and the last dataset was lost. The tare data collected before the third passage was normalized using the pre-welding tare and used as a last data point. Axial and shear strain are observed to peak and then quickly decrease. The azimuthal strain peak is less pronounced, and the strain is overall increasing as it is expected. The average axial and azimuthal strain values are observed to decrease overnight between the first and the second passage suggesting that it takes some time for the SS shell to thermalize and adjust. The final value reached after the second passage are very close to the one observed after the first ones. This behavior suggests that the last strain data point values used to calculate the shell stress are a good representation of the actual strain on the shell at the end of the third passage. We measured an average axial strain of = -14.3207 µε, an Azimuthal strain of = 78.7379 µε and a Shear strain of = -39.1052 µε.

The average stress was calculated using 200 GPa for the stainless-steel Young modulus [15]. We found that the average azimuthal stress generated on the SS outer shell during welding is around 15.7 MPa, which is very close to what has been computed. However, this measured stress value is conservative with respect to the computational data. The effect of bending cannot be quantified. Moreover, the computed SS-shell stress value (15.5 MPa) is the average between the outer (21 MPa) and inner (10 MPa) SS shell stress values [6]. In conclusion, the SS-shell welding has been performed according to requirements.

This test demonstrated that the use of a sensor grid is more effective than the use of traditional strain gauges to validate strain during welding. Indeed, it was not only possible to evaluate the average strain during welding but also to obtain a strain map of a very large area providing a more precise and realistic information about the strain variation on the SS vessel. A stress validation plan has been established for the future pre-series and series LMQXFA cold masses that are going to be fabricated. This plan consists in installing two fiber grids on each cold mass: a 20 m sensor on the top shell quadrant that covers a 1400mmx 500 mm area and a 10 m sensor on the bottom shell quadrant (700x500 mm). This is because blocks from the welding tooling are in contact with the bottom shell.

## III. CONCLUSION

A novel way to use optical sensors, based on distributed fiber optic, was developed to obtain a strain map over a large area of the stainless-steel shell of Nb₃Sn magnets. A fiber grid was designed and successfully tested to measure the strain variations of the first LMQXFA SS shell during welding. The measured stress allows to infer compliance with specifications and validate the welding process requirements as well as the shell design. This technique presents several advantages compared to strain gauges. Indeed, a 100 thousand times larger area was covered with a single data acquisition, making the measured values more representative in describing the actual strain on the SS vessel.

This novel way to use distributed fiber is very promising and could be employed to obtain for the first time a strain map of superconducting Nb3Sn coil during magnet cooldown, powering and training.